\documentclass[prb,preprint]{revtex4-1}
\usepackage{amsmath}  % needed for \tfrac, \bmatrix, etc.
\usepackage{amsfonts} % needed for bold Greek, Fraktur, and blackboard bold
\usepackage{graphicx} % needed for figures

\renewcommand{\vec}[1]{\mbox{\boldmath $#1$}}
\newcommand*\Del{\mathrm{\Delta}}                 % Roman Delta
\newcommand{\uvec}[1]{\hat{\vec #1}}
\def\arcsec{\hbox{$^{\prime\prime}$}}
\begin{document}

% Be sure to use the \title, \author, \affiliation, and \abstract macros
% to format your title page.  Don't use lower-level macros to manually
% adjust the fonts and centering.

\title{Photon in a cavity---a Gedankenexperiment}
% In a long title you can use \\ to force a line break at a certain location.

\author{Klaus Wilhelm}
\email{wilhelm@mps.mpg.de} % optional
%\altaffiliation[permanent address: ]{101 Main Street,
%  Anytown, USA} % optional second address
% If there were a second author at the same address, we would put another
% \author{} statement here.  Don't combine multiple authors in a single
% \author statement.
\affiliation{Max-Planck-Institut f\"ur Son\-nen\-sy\-stem\-for\-schung
(MPS), 37191 Katlenburg-Lindau, Germany}

\author{Bhola N. Dwivedi}
\email{bholadwivedi@gmail.com}
\affiliation{Dept. of Applied Physics, Indian Institute of Technology
(Banaras Hindu University), Varanasi-221005, India}

% See the REVTeX documentation for more examples of author and affiliation

\date{\today}

%%%%%%%%%%%%%%%%%%%%%%%%%%%%%%%%%%%%%%%%%%%%%%%%%%%%%%%%%%%%%%%%%%%%%%%%%%%%%%
\begin{abstract}
The inertial and gravitational mass of electromagnetic radiation
(i.e., a photon distribution) in a cavity with reflecting walls has been
treated by many authors for over a century. After many contending discussions,
a consensus has emerged that the mass of such a
photon distribution is equal to its total energy divided by the square of the
speed of light. Nevertheless, questions remain unsettled on the interaction of
the photons with the walls of the box. In order to understand some of the
details of this interaction, a simple case of a single photon
with an energy~$E_\nu = h\,\nu$ bouncing
up and down in a static cavity with perfectly reflecting walls
in a constant gravitational field~$\vec{g}$, constant in space and time,
is studied and its contribution to the weight of the box is determined as a
temporal average.
\end{abstract}
% AJP requires an abstract for all regular article submissions.
% Abstracts are optional for submissions to the "Notes and Discussions"
% section.
%%%%%%%%%%%%%%%%%%%%%%%%%%%%%%%%%%%%%%%%%%%%%%%%%%%%%%%%%%%%%%%%%%%%%%%%%%%%%%
\maketitle % title page is now complete

%%%%%%%%%%%%%%%%%%%%%%%%%%%%%%%%%%%%%%%%%%%%%%%%%%%%%%%%%%%%%%%%%%%%%%%%%%%%%%
\section{Introduction} % Section titles are automatically converted to
% all-caps.
% Section numbering is automatic.
\label{s.introd}
%% Sect. I

Massive particles and electromagnetic radiation (photons) in a box
have been considered by many authors.
\cite{Poi00,Abr04,Has05,Ein05,Ein06,Ein08,Lau20,BasSch,Fee60,Ant76,Kol95,Riv95,RueHai}
Most of them have discussed the inertia of an
empty box in comparison with a box filled with a gas or radiation.

\begin{figure}[ht!]
%% FIG. 1
% The bracketed code determines the figure's placement:  "h" stands for
% "here", telling LaTeX to put the figure as close to the current location
% as possible.  The ! overrides LaTeX's tendency to try to find a location
% that it thinks is better.  But don't agonize over the exact figure placement
% in your submitted manuscript.  For your initial submission, just make sure
% each figure is reasonably close to where it's first referenced.
\centering
\includegraphics[width=12.5cm]{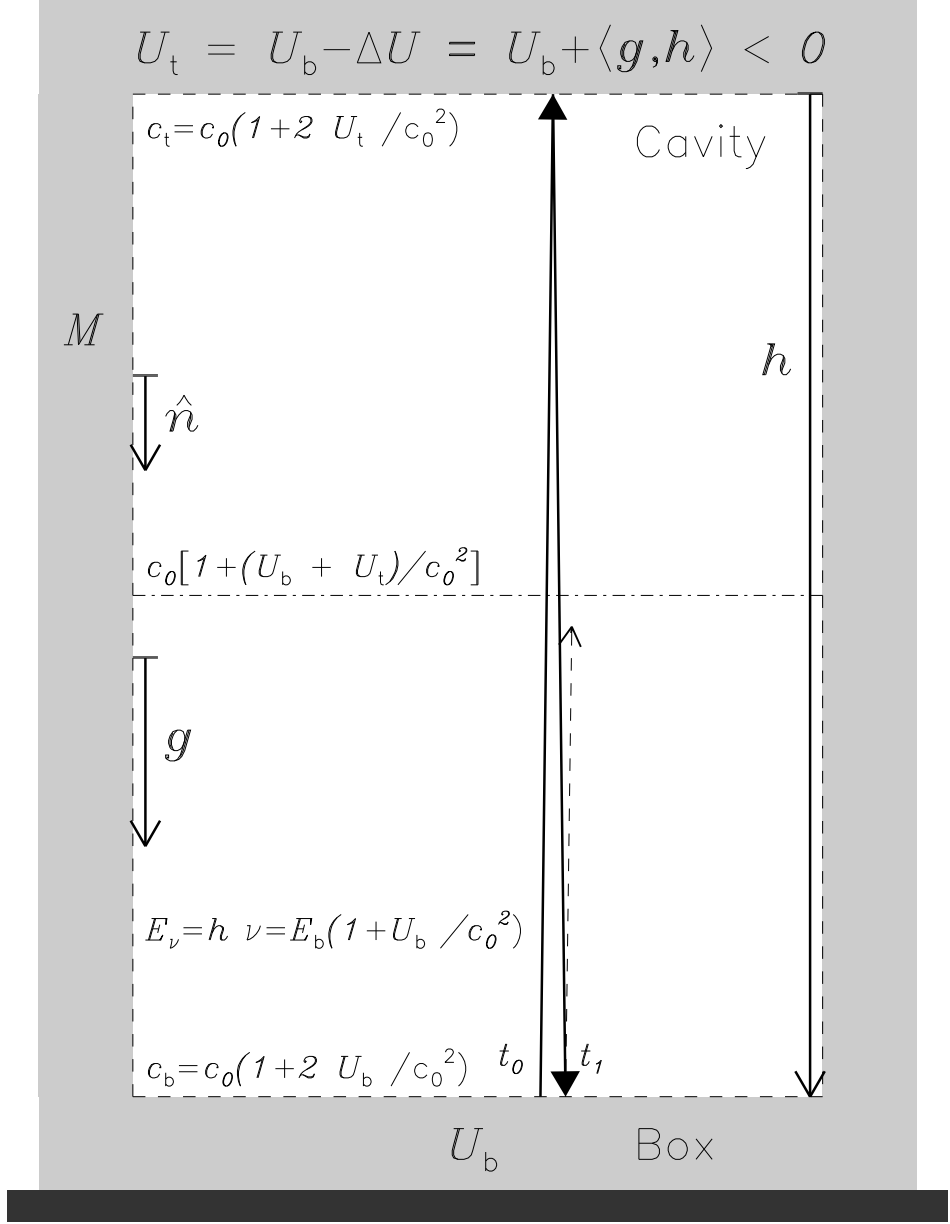}
\caption{Box of height~$h =||\vec{h}||$ with mass~$M$ and reflecting inner
walls on a weighing scale (indicated by the black bar)
to determine its weight in a constant gravitational field~$\vec{g} =
g\,\uvec{n}$. The gravitational potentials at the bottom and top are
$U_{\rm b}$ and $U_{\rm t}$, respectively. One cycle with a
period of $T = t_1 - t_0$ and the continuation into the next cycle
(dashed arrow) are schematically
shown for a photon with an energy~$E_\nu = h\,\nu$ bouncing (nearly)
vertically up and down.
The values of the speed of light~$c_{\rm b}$ and
$c_{\rm t}$ in the cavity
corresponding to the potentials~$U_{\rm b}$ and $U_{\rm t}$ are given at the
bottom and the top as well as for the center as a mean speed of
$c_0\,(1 + 2\,\overline{U}/c^2_0)$.}
\label{f.Box_1}
\end{figure}

In the presence of a constant gravitational
field~$\vec{g}$ (pointing downwards in Fig.~\ref{f.Box_1}), the
effect on the weight of the box is another topic that has been studied. In
very general terms, this problem has been treated in
Refs.~\onlinecite{Lau20,Ein08} with the conclusion that, in a closed system
in equilibrium, all types of energy~$E_n$ contribute to the mass according to
%
%% Eq. 1
\begin{equation}
\Del M =\frac{\sum E_n}{c^2_0}~,
\label{Energies}
\end{equation}
where~$M$ refers to the passive gravitational mass.

For a gas and even for a single massive particle, this can easily be verified
with the help of the energy and momentum conservation laws as a temporal
average. For radiation the situation has been debated over the
years.\cite[cf.][]{Ant76,Hec11} Kolbenstvedt in Ref.~\onlinecite{Kol95}
studied photons in a uniformly accelerated cavity and found a mass
contribution
\begin{quote}
[...] in agreement with Einstein's mass-energy formula.
\end{quote}
It might, therefore, be instructive to
describe the problem with the help of a Gedankenexperiment
in the simple case of a photon bouncing up and down in a
cavity of a box with perfectly reflecting inner walls \emph{at rest}
in a constant gravitational field. The height~$\vec{h}$ is measured as
fall height in the field direction, which will be
indicated by the unit vector $\uvec{n}$ in equations
and figures. The mass of the box, $M$,
includes the mass of the walls and the equivalent mass of any unavoidable
energy content, such as thermal radiation, except the test
photon.
%
% When an AJP manuscript is conditionally accepted for publication, we
% will request an editable copy of the manuscript in which the figures
% themselves have been removed, and all figure captions have been moved
% to the end.  You should then comment-out the \includegraphics line
% and move the entire \begin{figure} ... \end{figure} block to the end
% of the source file, just before \end{document}.

\section{Definition of gravitational potentials} %%%%%%%%%%%%%%%%%%%%%%%%%%%%
\label{s_pot}
%% Sect. II

The gravitational potentials will be defined as $U_{\rm t}$ at the
top of the cavity and as $U_{\rm b}$ at the bottom with
%
%% Eq. 2
\begin{equation}
U_{\rm b} - U_{\rm t} = \Del U  = - \vec{g}\cdot\vec{h} < 0
\label{potential}
\end{equation}
and the relations
%
%% Eq. 3
\begin{equation}
- c^2_0 \ll {U_{\rm b} < \overline{U} < U_{\rm t} < 0}~,
\label{Potential}
\end{equation}
where~$c_0$ is the speed of light in vacuum without a gravitational field
and $\overline{U} = (U_{\rm b} + U_{\rm t})/2$.
Under the weak-field condition, so defined, the speed of light measured
on the coordinate or world time scale is
\cite{AshAll,Sch60,Gui97,Oku00,Str00}
%
%% Eq. 4
\begin{equation}
c(U) = c_0\left(1 + \frac{2\,U}{c^2_0}\right) ~.
\label{Speed}
\end{equation}

Einstein originally derived from the equivalence principle
%
%% Eq. 5
\begin{equation}
c = c_0\left(1 + \frac{\phi}{c^2}\right)
\label{Wrong_Speed}
\end{equation}
with $\phi$ as symbol for the gravitational potential~$U$ used
here.\cite{Ein11}
Application of Huygens' principle then led him to expect a deflection of
0.83\arcsec~for a Sun-grazing light beam. A value of 0.84\arcsec~had been
obtained by Soldner in 1801.\cite{Sol01} However, in 1916 Einstein
predicted
a deflection of 1.7\arcsec~based on his general theory of
relativity.\cite{Ein16} This was first verified during a solar eclipse in
1919, and since then values between 1.75\arcsec~ and 2.0\arcsec~have been
found in many observational studies.\cite{Dysetal,Mik59,Shaetal}
Consequently, it can be concluded that observations are not consistent with
Eq.~(\ref{Wrong_Speed}), but are\,---\,within the uncertainty margins\,---\,in
agreement with Eq.~(\ref{Speed}).

The relativistic energy ($E$) and momentum ($\vec{p}_0$) equation for a free
body with mass~$m$ in vacuum is
%
%% Eq. 6
\begin{equation}
E^2 = m^2\,c^4_0 + ||\vec{p_0}||^2\,c^2_0 ~.
\label{En_mom_eq}
\end{equation}
It reduces for a massless particle to
%
%% Eq. 7
\begin{equation}
E = p_0\,c_0
\label{massless}
\end{equation}
with $p_0 = ||\vec{p}_0||$.\cite{Ein17,Dir36,Oku89}
In a static gravitational field the energy of a photon
%
%% Eq. 8
\begin{equation}
E_\nu = h\,\nu = p(U)\,c(U)
\label{constant}
\end{equation}
measured in the coordinate time system is constant.\cite{Oku00}
Its speed, however, varies according to Eq.~(\ref{Speed}), whilst the
momentum changes inversely to this speed.

\section{Photon reflection scenarios} %%%%%%%%%%%%%%%%%%%%%%%%%%%%%%%%%%%%%%%
\label{s_scenarios}
%% Sect. III

At least three different scenarios can be conceived to describe the situation
inside the box:

\begin{enumerate}
\item A naive application of Eqs.~(\ref{Speed}) and (\ref{constant}), i.e.,
direct reflections at the walls without any further interaction,
leads to the result that the photon contributes
%
%% Eq. 9
\begin{equation}
\Del m \approx
2\,\frac{E_\nu}{c^2_0}\left(1 +
%\frac{U_{\rm b}}{c^2_0} + \frac{U_{\rm t}}{c^2_0}\right)
\frac{2\,\overline{U}}{c^2_0}\right)
\label{Contrib_1}
\end{equation}
to the mass of the box, which is nearly a factor of two higher than
expected from
Eq.~(\ref{Energies}). To show this, let us consider the oppositely directed
momentum transfers during photon reflections at the bottom and the ceiling of
the cavity. The values are twice
%
%% Eq. 10
\begin{equation}
%\vec{p}^{\rm C}_{\rm b} =
\frac{E_\nu}{c_{\rm b}}\,\uvec{n}
\approx
\vec{p}_\nu\left(1 - 2\,\frac{U_{\rm b}}{c^2_0}\right)
\label{momentum_b}
\end{equation}
and
%
%% Eq. 11
\begin{equation}
%- \vec{p}^{\rm C}_{\rm t} =
- \frac{E_\nu}{c_{\rm t}}\,\uvec{n}
\approx - \vec{p}_\nu\left(1 - 2\,\frac{U_{\rm t}}{c^2_0}\right) ~,
\label{momentum_t}
\end{equation}
where $c_{\rm b}$ and $c_{\rm t}$ are determined from
Eq.~(\ref{Speed}) taking into account the conditions in~(\ref{Potential})
to justify the approximations, i.e., neglecting
orders equal or higher than $(U/c^2_0)^2$ against unity.
The momentum $\vec{p}_\nu =(E_\nu/c_0)\,\uvec{n}$ of a photon with
energy~$E_\nu = h\,\nu$ at $U_0 = 0$ in vacuum has been introduced.
A complete cycle lasts for
%
%% Eq. 12
\begin{equation}
T = t_1 - t_0 =
\frac{2\,h}{c_0\,(1 + 2\,\overline{U}/c^2_0)} ~,
\label{Time}
\end{equation}
i.e., $2\,h$ divided by the mean speed. Within this time interval, the total
momentum transfer of
%
%% Eq. 13
\begin{equation}
\Del \vec{P}^{\rm C} =
2\,\left(\frac{E_\nu}{c_{\rm b}} - \frac{E_\nu}{c_{\rm t}}\right)\,\uvec{n}
%2\,(\vec{p}^{\rm C}_{\rm b} - \vec{p}^{\rm C}_{\rm t})
\approx 4\,\vec{p}_\nu\,\frac{\vec{g}\cdot\vec{h}}{c^2_0} ~,
\label{Delta_p}
\end{equation}
is obtained from Eqs.~(\ref{potential}), (\ref{momentum_b}) and
(\ref{momentum_t}). A momentum vector with upper index~C
refers in this and later equations to a momentum inside the cavity.
Division by~$T$ of Eq.~(\ref{Time}) gives a mean force of
%
%% Eq. 14
\begin{equation}
\overline{\vec{F}_1} \approx
2\,\frac{E_\nu}{c^2_0}\left(1 +
\frac{2\,\overline{U}}{c^2_0}\right)\vec{g} ~,
\label{force}
\end{equation}
confirming Eq.~(\ref{Contrib_1}),
which has been found to be  inconsistent with Eq.~(\ref{Energies}).

\item
The assumption that the reflections occur at the walls of the box without
further effects might not be correct. Consequently, the next scenario is based
on an intermediate storage of the energy~$E_\nu$, i.e., as elastic energy,
and a transfer of a momentum calculated under the assumption that the
speeds~$c_{\rm b}$ and $c_{\rm t}$ valid in the cavity
at $U_{\rm b}$ and $U_{\rm t}$, respectively, are also applicable
for the complete reflection process.
This gives a momentum transfer of
%
%% Eq. 15
\begin{equation}
2\,\vec{p}_\nu\left(1 - \frac{2\,U_{\rm b}}{c^2_0}\right)
\label{Mom_bot_2}
\end{equation}
during the absorption and emission at the bottom and
%
%% Eq. 16
\begin{equation}
- 2\,\vec{p}_\nu\left(1 - \frac{2\,U_{\rm t}}{c^2_0}\right)
\label{Mom_top_2}
\end{equation}
at the top in opposite directions. Comparison with
Eqs.~(\ref{momentum_b}) and (\ref{momentum_t}) shows that
the resulting
force~$\overline{\vec{F}_2}$ after application of Eqs.~(\ref{Time}) to
(\ref{force})
equals $\overline{\vec{F}_1}$ and leads to the same surprising result.

%
%% ONE-COLUMN FIGURE
% FIG. 2
\begin{figure}
\centering
\includegraphics[width=10cm]{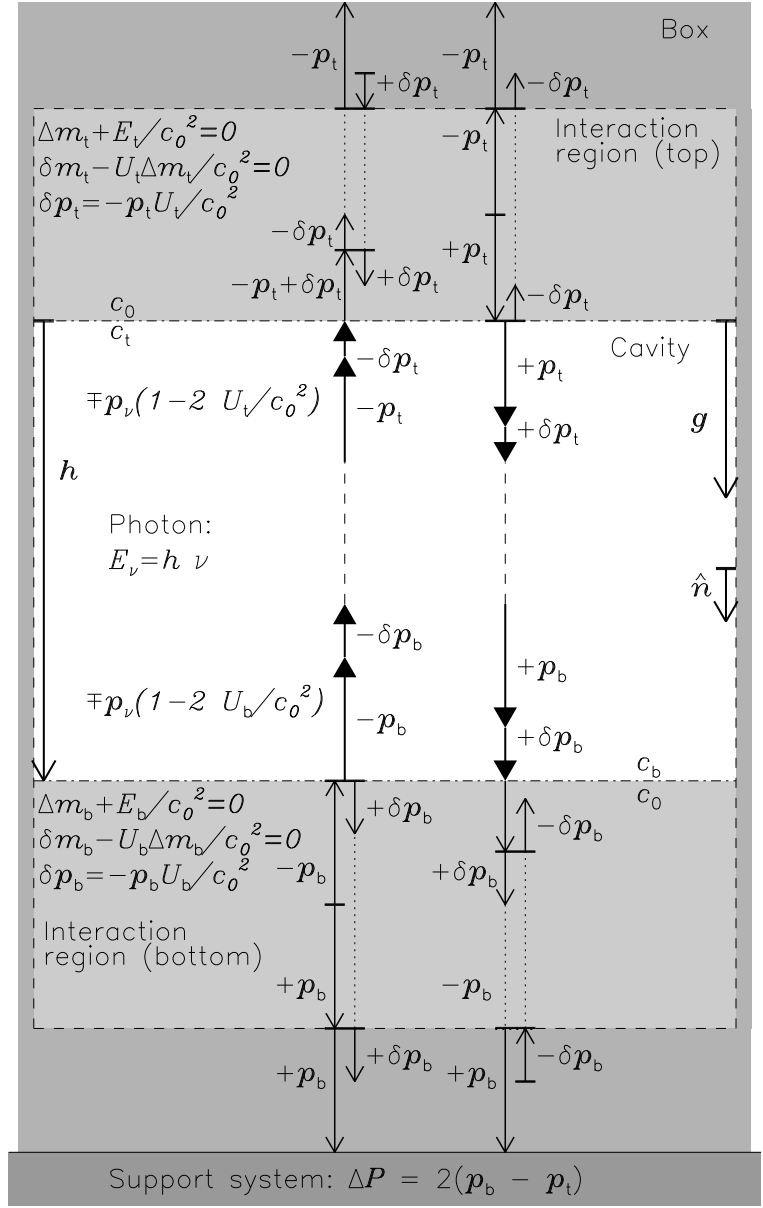}
\caption{The box is shown on a weighing scale as support system. The bottom
and the top of the cavity are complemented by ``interaction regions''.
Energy release and photon emission processes occur in these regions as
required by \emph{energy and momentum conservation}. The initial energy
release of~$E_{\rm b} = \Del m_{\rm b}\,c^2_0$ happens in the bottom
interaction region accompanied by a momentum transfer to the box and the
support system of~$+\vec{p}_{\rm b}$. The conversion of~$\Del m_{\rm b}$ into
energy at~$U_{\rm b}$ is subject to an increase of the potential energy of the
bottom interaction region by~$- U_{\rm b}\,\Del m_{\rm b}$ taken
from~$E_{\rm b}$. The related differential momentum
contribution~$+ \delta\vec{p}_{\rm b}$ acts on the support system. It will be
compensated by the opposite effect of the following energy-to-mass conversion
at~$U_{\rm b}$ after one period~$T$. The corresponding steps at~$U_{\rm t}$
are described in the text.
\label{f.Box_2}}
\end{figure}

\item
The previous concept was based on the assumption that the reflections actually
occur in regions of the wall, where the effective speed of light has been
modified by the local gravitational potential. However, the process of a
photon reflection has to be accomplished by interactions with electrons in
the walls. The huge ratio of the electrostatic to the gravitational forces
between elementary particles makes it unlikely that such an interaction is
directly influenced significantly by weak fields of gravity. However, the
\emph{photon absorption} and \emph{emission} will be affected as is evident
from the gravitational redshift. This redshift---since its prediction in 1908
by Einstein in Ref.~\onlinecite{Ein08}---has
been theoretically and experimentally studied in many
investigations\cite{Lau20,StJ28,Oku00,PouReb,Sch60,Craetal,Hayetal,BlaRod,KraLue,
Bra63,PouSni,Sni72,Wil74,Okuetal,Wol10,SinSam} resulting in a
quantitative validation of Einstein's statement in Ref.~\onlinecite{Ein11}:
\begin{quote}
[...] m\"ussen also die Spektrallinien des Sonnenlichts gegen\"uber den
entsprechenden Spektrallinien irdischer Lichtquellen etwas nach dem Rot
verschoben sein, und zwar um den relativen Betrag
%
%% Eq. 17
\begin{equation}
\frac{\nu_0 - \nu}{\nu_0} = \frac{- \phi}{c^2} = 2 \cdot 10^{-6} ~.
\label{Ein_red}
\end{equation}
([...] spectral lines of light from the Sun must consequently be shifted
somewhat towards the red with respect to the corresponding spectral lines
of terrestrial light sources, namely by the relative value of [...]
$2 \cdot 10^{-6}$.)
\end{quote}
The (negative) difference of the
gravitational potentials between the Sun and the Earth is denoted
by $\phi$ in Eq.~(\ref{Ein_red}).

Einstein's early suggestion was that the
transition of an atom is an intra-atomic process, i.e. it is not dependent
on the gravitational potential:
\begin{quote}
Da der einer Spektral\-linie ent\-spre\-chen\-de
Schwin\-gungs\-vorgang wohl als
ein intraatomischer Vorgang zu betrachten ist, dessen Frequenz durch das Ion
allein bestimmt ist, so k\"onnen wir ein solches Ion als eine Uhr von
be\-stimm\-ter Frequenzzahl $\nu_0$ ansehen.\cite{Ein08}

(Since the oscillation process corresponding to a spectral line
probably can be envisioned as an intra-atomic process, the frequency of which
is determined by the ion alone, we can consider such an ion as a clock with
a distinct frequency~$\nu_0$.)
\end{quote}
This means that the energy~$E_{\rm b}$ initially released at the
gravitational potential~$U_{\rm b}$ by an elementary process equals the
energy~$E_0$
released by the \emph{same} process at the potential $U_0 = 0$. In both cases
the process will be accompanied by a momentum pair
%
%% Eq. 18
\begin{equation}
\pm \vec{p}_{\rm b} = \pm \frac{E_{\rm b}}{c_0}\,\uvec{n} =
\pm \frac{E_0}{c_0}\,\uvec{n}~.
\label{momentum}
\end{equation}
It has, however, to be noted in this
context that Einstein later concluded:
\begin{quote}
Die Uhr l\"auft also langsamer, wenn sie in der N\"ahe ponderabler
Massen aufgestellt ist. Es folgt daraus, da{\ss} die Spektralinien von der
Oberfl\"ache gro{\ss}er Sterne zu uns gelangenden Lichtes nach dem roten
Spektralende verschoben erscheinen m\"ussen.\cite{Ein16}

(The clock is thus delayed, if it is placed near ponderable masses.
Consequently, it follows that the spectral lines of light reaching us from the
surface of large stars must be shifted towards the red end of the spectrum.)
\end{quote}
Both statements can be reconciled by postulating that the redshift occurs
during the actual emission process and realizing that energy trapped
in a closed
system has to be treated differently from propagating radiation energy.

In addition, the importance of the momentum transfer during the absorption
or emission of radiation was emphasized by
Einstein:
\begin{quote}
Bewirkt ein Strahlenb\"undel, da{\ss} ein von ihm getroffenes Molek\"ul die
Ener\-gie\-menge~$h\,\nu$ in Form von Strahlung durch einen Elementarproze{\ss}
auf\-nimmt oder abgibt (Einstrahlung), so wird stets der
Impuls~$\frac{\displaystyle {h\,\nu}}{\displaystyle c}$ auf das Molek\"ul
\"ubertragen, und zwar bei der
Energieaufnahme in der Fortpflan\-zungs\-richtung des B\"undels, bei der
Energieabgabe in der entgegengesetzten Richtung. [...].

Aber im allgemeinen begn\"ugt man sich mit der Betrachtung des
E\,n\,e\,r\,g\,i\,e-Austausches, ohne den I\,m\,p\,u\,l\,s-Austausch
zu ber\"uck\-sich\-tigen.\cite{Ein17}

(A beam of light that
induces a molecule to absorb or deliver the energy~$h\,\nu$
as radiation by an elementary process (irradiation) will always transfer the
momentum~$\frac{\displaystyle {h\,\nu}}{\displaystyle c}$ to the molecule,
directed in the propagation direction of the beam for energy absorption, and
in the opposite direction for energy emission. [...].

However, in general one is satisfied with
the consideration of the e\,n\,e\,r\,g\,y exchange, without taking
the m\,o\,m\,e\,n\,t\,u\,m exchange into account.)
\end{quote}
The energy and momentum conservation principles lead to a
relationship between the energy~$E_{\rm b}$
and the emitted photon energy $E_\nu$ at $U_{\rm b}$ of
%
%% Eq. 19
\begin{eqnarray}
E_\nu = E_{\rm b} - ||\delta \vec{p}||\,c_0 =
||\vec{p}_{\rm b} - \delta \vec{p}||\,c_0  = \nonumber \\
||\vec{p}_{\rm b} + \delta \vec{p}_{\rm b}||\,c_{\rm b} =
||\vec{p}(U_{\rm b})||\,c_{\rm b}
\label{Energy}
\end{eqnarray}
by introducing a
differential momentum~$\delta \vec{p}_{\rm b}$ parallel to~$\vec{p}_{\rm b}$
(neglecting the very small recoil energy).
The evaluation gives, together with Eq.~(\ref{Speed}),
%
%% Eq. 20
\begin{equation}
\delta \vec{p}_{\rm b} = - \vec{p}_{\rm b}\,\frac{U_{\rm b}}{c^2_0}
\label{delta_p}
\end{equation}
and
%
%% Eq. 21
\begin{equation}
E_\nu =  E_{\rm b}\left(1 + \frac{U_{\rm b}}{c^2_0}\right) ~.
\label{Redshift}
\end{equation}
Such a scenario has recently been discussed in the context of the
gravitational redshift.\cite{WilDwi}

With the basic assumption that
the photon is reflected from both walls with the same
energy~$E_\nu$, the relation for the ceiling is:
%
%% Eq. 22
\begin{equation}
E_\nu = E_{\rm t}\left(1 + \frac{U_{\rm t}}{c^2_0}\right) ~.
\label{Redshift_t}
\end{equation}
The corresponding momentum transfers during absorption and emission
expected according to Eqs.~(\ref{momentum}), (\ref{Redshift}) and
(\ref{Redshift_t}) are
%
%% Eq. 23
\begin{equation}
2\,\vec{p}_{\rm b} \approx
2\,\frac{E_\nu}{c_0}\left(1 - \frac{U_{\rm b}}{c^2_0}\right)\uvec{n} =
2\,\vec{p}_\nu\left(1 - \frac{U_{\rm b}}{c^2_0}\right)
\label{Mom_bot_3}
\end{equation}
at the bottom and
%
%% Eq. 24
\begin{equation}
- 2\,\vec{p}_{\rm t} \approx
- 2\,\frac{E_\nu}{c_0}\left(1 - \frac{U_{\rm t}}{c^2_0}\right)\uvec{n} =
- 2\,\vec{p}_\nu\left(1 - \frac{U_{\rm t}}{c^2_0}\right)
\label{Mom_top_3}
\end{equation}
at the top. Note in this context
that the elementary processes at the bottom and the top must have slightly
different energy levels for constant $E_\nu$.
Comparison of the momentum values of Eqs.~(\ref{Mom_bot_3}) and
(\ref{Mom_top_3}) with those of Eqs.~(\ref{momentum_b}) and (\ref{momentum_t})
shows that the
force~$\overline{\vec{F}_3}$ obtained in analogy to Eq.~(\ref{Delta_p})
equals $\overline{\vec{F}_1}/2$ and thus gives a value expected from
Eq.~(\ref{Energies}).

This encouraging result will now be analysed in detail.
The initial energy release is assumed in the
``Interaction region (Bottom)'' at the potential~$U_{\rm b}$
in Fig.~\ref{f.Box_2} on the left according to
%
%% Eq. 25
\begin{equation}
E_{\rm b} + \Del m_{\rm b}\,c^2_0 = 0
\label{Release}
\end{equation}
accompanied by a momentum pair of~$\pm \vec{p}_{\rm b}$ as mentioned above.
The exact release process is of no importance for the present discussion, but
the initial release must be controlled by the elementary process alone. A
speed of $c_{\rm b}$ at the bottom of the cavity, together with energy and
momentum conservation laws, then requires that a photon can only be emitted
with an energy~$E_\nu$ given by Eqs.~(\ref{Energy}) and (\ref{Redshift})
(the rest energy of the mass~$\Del m_{\rm b}$ at the gravitational
potential~$U_{\rm b}$) and a momentum in the upward direction of
%
%% Eq. 26
\begin{equation}
- \vec{p}^{\rm C}_{\rm b} =
- \vec{p}_{\rm b} - \delta \vec{p}_{\rm b}
\approx - \vec{p}_\nu\left(1 - \frac{2\,U_{\rm b}}{c^2_0}\right) ~,
\label{mom_3}
\end{equation}
where we have used Eqs.~(\ref{delta_p}) and (\ref{Mom_bot_3}).
The energy difference, corresponding to the potential energy of a
mass~$\Del m_{\rm b}$ at~$U_0$ relative to~$U_{\rm b}$,
%
%% Eq. 27
\begin{equation}
E_{\rm b} - E_\nu = - U_{\rm b}\,\Del m_{\rm b}
\label{Difference}
\end{equation}
will be transferred to the box.
This process is accompanied by a corresponding
differential momentum transfer of~$+ \delta\vec{p}_{\rm b}$ and
by a mass increase of
%
%% Eq. 28
\begin{equation}
\delta m_{\rm b} = - \frac{U_{\rm b}\,\Del m_{\rm b}}{c^2_0} ~.
\label{d_m_b}
\end{equation}
The momentum~$+\vec{p}_{\rm b}$ from the initial release and
$+ \delta \vec{p}_{\rm b}$ will thus be acting on  the support system.

The reflections at the top and bottom have to be considered in several
steps\,---\,illustrated in detail in Fig.~2. :
\end{enumerate}
\begin{itemize}
\item
The photon $E_\nu$ arrives at the top with $c_{\rm t}$,
obtained from Eq.~(\ref{Speed}), and a momentum of
%
%% Eq. 29
\begin{eqnarray}
- \vec{p}^{\rm C}_{\rm t} =
- \vec{p}_{\rm b}\,\frac{(1 - U_{\rm b}/c^2_0)(1 + 2\,U_{\rm b}/c^2_0)}
{1 + 2\,U_{\rm t}/c^2_0}% \nonumber \\
\approx - \vec{p}_{\rm t} - \delta \vec{p}_{\rm t}
\approx - \vec{p}_\nu\left(1 - \frac{2\,U_{\rm t}}{c^2_0}\right)
\label{Moment_t}
\end{eqnarray}
derived from Eqs.~(\ref{constant}) and (\ref{Mom_top_3}) with
%
%% Eq. 30
\begin{equation}
\delta\vec{p}_{\rm t} = - \vec{p}_{\rm t}\,\frac{U_{\rm t}}{c^2_0} ~.
\label{delta_p_t}
\end{equation}
\item
The change of the speed from~$c_{\rm t}$ to $c_0$ in the interaction region
will entail
a change of the momentum in
Eq.~(\ref{Moment_t}) to a new value, which can be written in our
approximation as $-\vec{p}_{\rm t} + \delta \vec{p}_{\rm t}$.
\item
The energy-to-mass conversion at the potential~$U_{\rm t}$ according to
%
%% Eq. 31
\begin{eqnarray}
\Del m_{\rm t} + \frac{E_{\rm t}}{c^2_0} = 0
\label{Release_t}
\end{eqnarray}
in the upper
interaction region can only be accomplished by adding the potential energy
term
%
%% Eq. 32
\begin{equation}
E_{\rm t} - E_\nu = - U_{\rm t}\,\Del m_{\rm t}
\label{Difference_t}
\end{equation}
and a momentum of $- \delta\vec{p}_{\rm t}$, which will be provided by the
conversion of~$\delta m_{\rm t}$ into energy. The
momentum~$+ \delta p_{\rm t}$ of the momentum pair will act on the box.
In total a momentum
of~$-\vec{p}_{\rm t} + \delta \vec{p}_{\rm t}$ has thus to be taken up by the
box.

These processes can be seen as
the reversed actions performed by
Eqs.~(\ref{Release}) to (\ref{d_m_b}),
but now at $U_{\rm t}$.

\item
The return trip essentially occurs in the reverse order as shown on the
right side of Fig.~\ref{f.Box_2}. The energy release of~$E_{\rm t}$
will be accompanied by a momentum pair~$\pm \vec{p}_{\rm t}$.
The photon will be emitted in the downward direction with a momentum of
%
%% Eq. 33
\begin{equation}
\vec{p}^{\rm C}_{\rm t} =
+ \vec{p}_{\rm t} + \delta\vec{p}_{\rm t}
\approx + \vec{p}_\nu\left(1 - \frac{2\,U_{\rm t}}{c^2_0}\right) ~,
\label{mom_4}
\end{equation}
cf., Eqs.~(\ref{Mom_top_3}) and (\ref{Moment_t}).
The energy difference $E_{\rm t} - E_\nu$ restores the potential energy
of Eq.~(\ref{Difference_t}).
A momentum of~$- \vec{p}_{\rm t} - \delta \vec{p}_{\rm t}$ will be transferred
to the box in analogy to the processes at $U_{\rm b}$.

\item
The photon momentum at the bottom of the cavity will be
%
%% Eq. 34
\begin{eqnarray}
\vec{p}^{\rm C}_{\rm b} =
\vec{p}_{\rm t}\,\frac{(1 - U_{\rm t}/c^2_0)(1 +2\,U_{\rm t}/c^2_0)}
{1 + 2\,U_{\rm b}/c^2_0}% \nonumber \\
\approx +\vec{p}_{\rm b} + \delta \vec{p}_{\rm b}
\approx + \vec{p}_\nu\left(1 -\frac{2\,U_{\rm b}}{c^2_0}\right)
\label{Moment_p}
\end{eqnarray}
as expected from Eq.~(\ref{mom_3}).
\item
In analogy to the situation at~$U_{\rm t}$,
we find a momentum of~$+ \vec{p}_{\rm p} - \delta \vec{p}_{\rm b}$ that has
to be transferred to the support system.
\item
The total external momentum thus is with Eqs.~(\ref{potential}),
(\ref{Potential}), (\ref{momentum}), (\ref{Redshift}) and (\ref{Redshift_t}):
%
%% Eq. 35
\begin{eqnarray}
\Del\vec{P} = (\vec{p}_{\rm b} + \delta \vec{p}_{\rm b})
- (\vec{p}_{\rm t} - \delta \vec{p}_{\rm t})
- (\vec{p}_{\rm t} + \delta \vec{p}_{\rm t})
+  (\vec{p}_{\rm b} - \delta \vec{p}_{\rm b}) =
2\,(\vec{p}_{\rm b} - \vec{p}_{\rm t}) \nonumber \\
= 2\,\left(\frac{E_{\rm b}}{c_0} - \frac{E_{\rm t}}{c_0}\right) \approx
- 2\, \frac{E_\nu}{c_0}\,\frac{\Del U}{1 + 2\,\overline{U}/c^2_0} =
2\,\frac{E_\nu}{c_0}\,\frac{\vec{g}\cdot\vec{h}}{1 + 2\,\overline{U}/c^2_0} ~.
\label{total}
\end{eqnarray}
Averaged over the time interval $T$ from Eq.~(\ref{Time}) for a full
cycle gives a mean force of
%
%% Eq. 37
\begin{equation}
\overline{\vec{F}_3} = \frac{\Del \vec{P}}{T}\approx
\frac{E_\nu}{c^2_0}\,\vec{g}
\label{correct_force}
\end{equation}
in agreement with Eq.~(\ref{Energies}).

A short comment is required on the intermediate mass storage processes
at the bottom and the top. Compared to $T$ the storage times are so small
that the contributions to the mean mass and thus the weight can be
neglected.
Since a factor of two was in question, there was also no need to complicate
the calculations even further by retaining
terms which are very small under the weak-field conditions assumed.

\end{itemize}

A multi-step process including Einstein's assumption of an intra-atomic
energy liberation thus leads to the correct result within our
approximations.
It involves an external box which cannot be completely rigid as it must be
able to enact the required gravitational energy and momentum transfers.
Another objection against a rigid box is the limited speed of any signal
transmission in the side walls.\citep[cf.][]{Ant76,Hec11}

\section{Discussion}%%%%%%%%%%%%%%%%%%%%%%%%%%%%%%%%%%%%%%%%%%%%%%%%%%%%%%%%%
\label{s.discuss}
%% Sect. IV

The assumption of a constant field~$\vec{g}$ in the environment of the
box---made in the interest of simplifying the calculations---calls for some
explanations. Einstein discussed in Ref.~\onlinecite{Ein12} a static
gravitational field without mass (,,[...] ein massenfreies statisches
Gravitationsfeld [...]'') and clarified in a footnote that such a field can be
thought of as generated by masses at infinity. Bondi in Ref.~\onlinecite{Bon86}
went even further in his scepticism against constant gravitational fields.
\emph{Only} the non-uniformity of the field would be observable. A field with
constant magnitude and direction should not be regarded as a field.

Indeed, there can be no perfectly homogeneous gravitational field.
Nevertheless very good approximations remote from the generating masses
exist between neighbouring potential surfaces, and the effects \emph{are}
observable. These remarks are of relevance for the considerations in the
previous section, as the calculations could have significantly been shortened
by setting $U_{\rm t} = 0$, but no realistic configuration would correspond to
such a definition. It can thus be concluded that Eq.~(\ref{potential}) should
be amended by $|\Del U| \ll |U_{\rm b}| \approx |U_{\rm t}|$ in order to
justify the assumption of a (nearly) constant field.

The present authors argued in Ref.~\onlinecite{WilDwi} that the interaction
of the liberated energy during an atomic transition with the kinetic
energy of the emitter and its momentum discussed by Fermi in relation to the
Doppler effect\cite{Fer32} has some resemblance with the gravitational
redshift, if the kinetic energy in the multi-step process leading to the
Doppler shift is replaced by the
potential energy. The same argument appears to be relevant for the photon
reflection process discussed here. The influence of gravity could indeed be
cancelled by the Doppler effect in the experiment of Pound and Rebka
(Ref.~\onlinecite{PouReb}) in such a way that the emission and absorption
energies of X-ray photons generated by the 14.4 keV transition of iron
(Fe$^{57}$) were the same in the source and the receiver positioned at
different heights in the gravitational field of the Earth.\cite{PouSni}

Although we have studied a single photon bouncing in a vacuum cavity, our
result might have some implications for the Abraham--Minkowski
controversy\cite{Min08,Abr10}
about the momentum of light in a medium with a refractive
index~$n = c_0/c > 1$. Modern expositions of this problem
have been presented, for instance, in Refs.~\onlinecite{ObuHeh,Bow05}.
The Abraham momentum is smaller in the medium than in vacuum, whereas the
Minkowski form gives a greater momentum inside the medium. In a recent paper,
\cite{Bar10} the dilemma could be resolved by identifying the
Abraham form with the kinetic momentum and the Minkowski formulation with the
canonical momentum. Both expressions are identical in
vacuum,\cite{Bow05} but this is only true in the absence of a gravitational
field. With such a field present, the Minkowski momentum is clearly
compatible with Eqs.~(\ref{Speed}) and (\ref{constant}) in the cavity.
In the interaction regions, the effective speed during
the momentum conversion processes appears to be the speed of light
in vacuum~$c_0$ not affected by the weak gravitational field, see
Eq.~(\ref{momentum}). Consequently, both Minkowski's and Abraham's
momentum calculations will lead to the same result.

\section{Conclusion}%%%%%%%%%%%%%%%%%%%%%%%%%%%%%%%%%%%%%%%%%%%%%%%%%%%%%%%%%
\label{s.concl}
%% Section 5

The temporally averaged increase of the passive gravitational mass and thus
the weight of a box containing a single photon bouncing up and down in a weak
gravitational field could be obtained in agreement with Eq.~(\ref{Energies})
by assuming interactions between the photon and the walls of the cavity
as well as momentum and energy conservation.

Preston formulated and wished some 130 years ago:
\begin{quote}
[...] let us work
towards the great generalization of the Unity of Matter and Energy.\cite{Pre83}
\end{quote}
Most of the work has been done by now, but it has also become quite
clear that the interactions between photons and matter in a
gravitational field are complicated as outlined by Bondi in his
article\cite{Bon97} entitled
``Why gravitation is not simple''.

%%%%%%%%%%%%%%%%%%%%%%%%%%%%%%%%%%%%%%%%%%%%%%%%%%%%%%%%%%%%%%%%%%%%%%%%%%%%%
%%%%%%%%%%%%%%%%%%%%%%%%%%%%%%%%%%%%%%%%%%%%%%%%%%%%%%%%%%%%%%%%%%%%%%%%%%%%%
%%%%%%%%%%%%%%%%%%%%%%%%%%%%%%%%%%%%%%%%%%%%%%%%%%%%%%%%%%%%%%%%%%%%%%%%%%%%%

\begin{acknowledgments}
 We thank the editor and an anonymous referee for constructive comments.
This research has made extensive use of the Astrophysics Data System (ADS).
\end{acknowledgments}

%%%%%%%%%%%%%%%%%%%%%%%%%%%%%%%%%%%%%%%%%%%%%%%%%%%%%%%%%%%%%%%%%%%%%%%%%%%%%

\end{document}